\documentstyle[aps,pre,multicol,epsfig]{revtex}

\newcommand{\be}{\begin{equation}}
\newcommand{\ee}{\end{equation}}
\newcommand{\bea}{\begin{eqnarray}}
\newcommand{\eea}{\end{eqnarray}}

\newcommand{\beet}{\begin{equation*}}
\newcommand{\eeet}{\end{equation*}}
\newcommand{\beaet}{\begin{eqnarray*}}
\newcommand{\eeaet}{\end{eqnarray*}}
\newcommand{\bfig}{\begin{figure}}
\newcommand{\efig}{\end{figure}}
\newcommand{\bc}{\begin{center}}
\newcommand{\ec}{\end{center}}

\begin{document}

\title{Jamming, Force Chains and Fragile Matter}

\author{M.~E. Cates$^1$, J.~P. Wittmer$^1$, J.-P. Bouchaud$^2$, P.
Claudin$^2$}
\address{
$^1$ Dept. of Physics and Astronomy, University of Edinburgh\\
JCMB King's Buildings, Mayfield Road, Edinburgh EH9 3JZ, UK.}
\address{$^2$ Service de Physique de l'Etat Condens\'e, CEA\\
Ormes des Merisiers, 91191 Gif-sur-Yvette Cedex, France.}

\maketitle

\begin{abstract}
We consider materials whose mechanical integrity
is the result of a jamming process. We argue that
such media are generically ``fragile": unable to support
certain types of incremental loading without plastic rearrangement.
Fragility is linked to the marginal stability of force chain
networks within the material. Such ideas may be relevant to
jammed colloids and poured sand. The crossover
from fragile (when particles are rigid) to elastoplastic behavior is 
explored.
\end{abstract}

\centerline{PACS numbers: 46.10.+z, 03.40.Kf}
\bigskip

\begin{multicols}{2}

\begin{figure}[tbh]
\centerline{\epsfysize=6cm
\epsfig{file=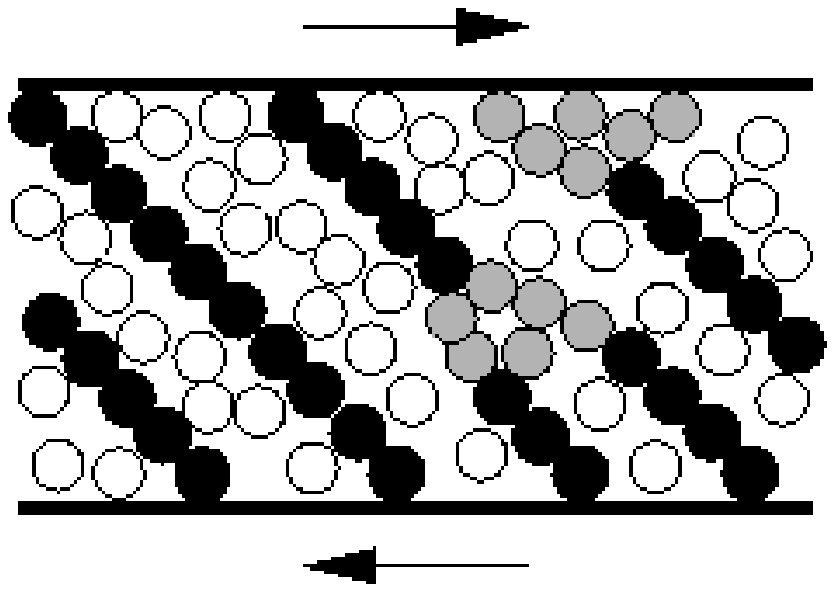,width=60mm,height=40mm,angle=0}}
\vspace*{0.4cm}
\centerline{\epsfysize=6cm
\epsfig{file=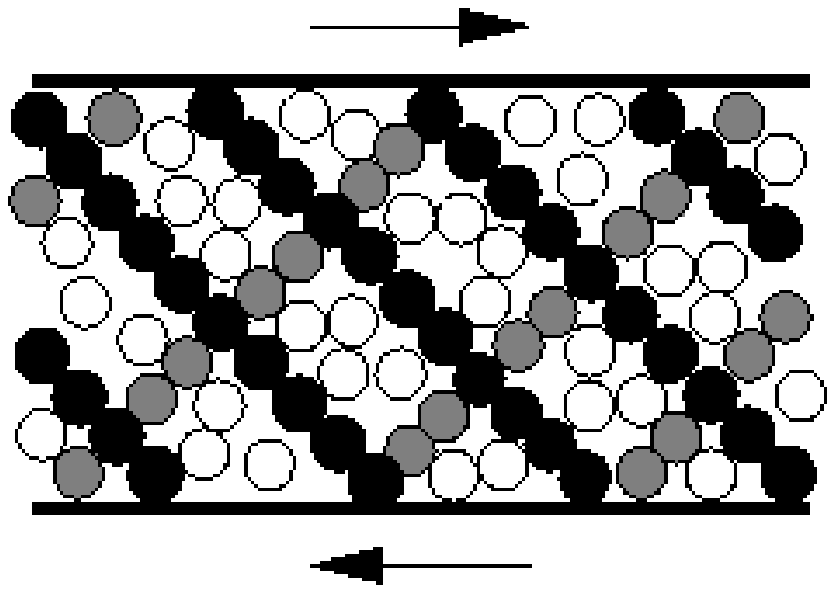,width=60mm,height=40mm,angle=0}}
{\small FIG.~1. (a) A jammed colloid (schematic).
Black: force chains; grey: other force-bearing particles; white:
spectators. (b) Idealized rectangular network of force chains.
\label{fig:jammed}}
\end{figure}

Consider a concentrated colloidal suspension of hard particles
under shear (Fig.~\ref{fig:jammed}1.(a)). Above a certain threshold
of stress, this system will jam \cite{farrmelrose}.
(To observe such an effect, stir a concentrated suspension
of corn-starch with a spoon.) In this Letter, we propose some
simple models of jammed systems like this, whose solidity stems
directly from the applied stress itself.
Such systems may be fundamentally different, in their mechanics,
from other classes of material, such as elastic or elastoplastic
solids.

In colloids, jamming apparently occurs because the particles form
``force chains" along the compressional direction \cite{farrmelrose}.
Even for spherical particles the lubrication films cannot prevent
contacts; once these arise, an array or network of force chains can, in
principle, support the shear stress indefinitely. (Brownian motion is
neglected here and below.) By this definition, the material is solid.

\begin{figure}
\centerline{\epsfysize=6cm
\epsfig{file=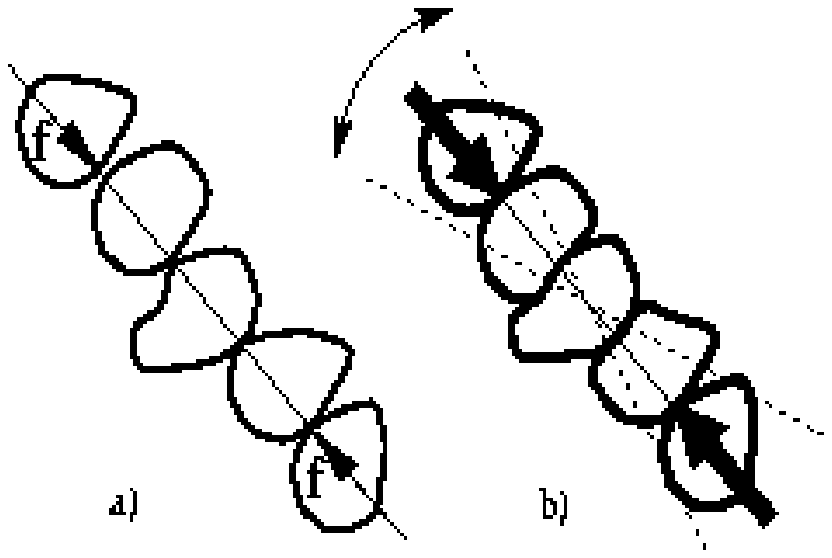,width=60mm,height=40mm,angle=0}}
{\small FIG.~2. (a) A force chain of hard particles (any shape) can
statically support only tangential compression. (Body forces
acting directly on these particles are neglected.)
(b) Finite deformability allows small transverse loads
to arise. \label{fig:paths} }
\end{figure}

A simple model of a force
chain assumes a linear string of rigid particles in point contact.
Crucially, this object can support only tangential loads \cite{Edwards}
(Fig.\ref{fig:paths}2.(a)): successive contacts must be
colinear, with the forces along the line of contacts, to
prevent torques on particles within the chain.
(Friction at the contacts does not obviate this requirement, nor
does particle asphericity.)

Let us model a jammed colloid by an assembly of such force chains,
characterized by a director
$\bf n$, in a sea of ``spectator" particles, and incompressible solvent.
(We ignore for the moment any
``collisions" between force chains or deflections caused by
weak interaction with the spectators.) In static equilibrium, with
no body forces acting, the compressive stress tensor is then
\be
\sigma_{ij} = P\delta_{ij} + \Lambda\, n_in_j
\label{nn}
\ee
where $P$ is an isotropic pressure, and $\Lambda$ a
compressive stress carried by the force chains.

Eq.(\ref{nn}) permits static equilibrium only so long as the applied
compression is
along $\bf n$; while this remains true, small, or even large,
incremental loads can be accommodated reversibly, by what is
(ultimately) an elastic mechanism. But the
material is certainly not an elastic solid, for if instead one
tries to shear the sample in a slightly different direction
(causing a rotation of the principal stress axes) static
equilibrium cannot be maintained without changing the director
$\bf n$.  And since $\bf n$ describes force chains
that pick their ways through a dense sea of spectator particles,
it cannot simply rotate; instead, the existing force chains must be
abandoned and new ones created with a slightly different orientation.
This entails
dissipative, plastic, reorganization, during which the system will
re-jam to support the new load.

The jammed colloid is an example of {\em fragile
matter}.
The medium can statically support applied
shear stresses (within some range), but does so by virtue of a
self-organized internal structure, whose mechanical properties have
evolved to support the load itself. Its incremental response can be
elastic only to {\em compatible} loads; incompatible loads (in this
case, those with a different compression axis), even if small, will
cause finite, plastic reorganizations. The system resembles a liquid
crystal, except that incompatible loads cause transient
rearrangement, not steady flow. An inability to elastically
support {\em some} infinitesimal loads is our definition of
``fragility". This extends naturally to small perturbations of other
types, such as changes in temperature, which can lead to
``static avalanches" of rearrangements \cite{staticavalanche}.

Jamming may lead {\em generically} to
fragile matter. If a system arrests as soon as it can support the
external load, its state is likely to be only marginally stable.
Incompatible perturbations induce rearrangements, leaving the
system in a newly jammed, equally fragile, state.
This scenario is related to suggestions that
rigidity emerges by successive buckling of force chains
(impeded by spectators) 
in glasses and granular matter \cite{alexander}.
It also resembles self-organized criticality ({\sc soc}) \cite{soc};
we return to this later 
(see also \cite{staticavalanche,cargese}). 
However, we focus first on simple
models of the fragile state in static equilibrium.

Consider again the (homogeneously) jammed colloid.
What body forces can it now support {\em without} plastic
rotation of the director? Various models are
possible. One is to assume that Eq.(\ref{nn}) continues to apply, with
$P({\bf r})$ and
$\Lambda({\bf r})$ now varying is space.
If $P$ is simply a fluid
pressure, a localized body force can be supported only if it acts
along $\bf n$. Thus (as in a bulk fluid) no static Green function
exists for a general body force. To support general loadings in three 
dimensions in fact requires
more than one orientation of force chain, perhaps forming a
network or skeleton \cite{dantu,thornton,kolymbas,radjai}.  A simple
model capable of describing this is:
\be
\sigma_{ij} = \Lambda_1\, n_in_j +
\Lambda_2\, m_im_j + \Lambda_3\, l_il_j\label{osl}
\ee
The directors ${\bf n},{\bf m},{\bf l}$ describe three
nonparallel populations of force chains and can also be seen
as {\it characteristics} along which the static Green function
propagates \cite{nature,bcc}. The $\Lambda$'s are
compressive pressures acting along these. Body forces cause
$\Lambda_{1,2,3}$ to vary in space.

Based on the above, we can distinguish
different levels of fragility, according to whether incompatible
loads include localized body forces (``bulk" fragility), or are
limited to forces acting at the boundary (``boundary" fragility).
(With Eq.(\ref{osl}), for example, a body force can always be
transmitted to the boundary but such boundary forces cannot then be
specified independently -- see below.) In disordered
systems one might also distinguish between macro-fragile responses
involving changes in the {\em mean} orientation of
force chains, and the micro-fragile responses of individual contacts.
Here we focus on macro-fragility; but if the medium is susceptible
to long-ranged `static avalanches'
\cite{staticavalanche} the distinction may be blurred.

Returning to the model of Eq.(\ref{osl}), the chosen values of the three
directors (two in 2-d) clearly should depend on how the system came to
be jammed (its ``construction history"). If it jammed in response to a
constant stress, switched on suddenly at some earlier time, one can
argue that the history is specified {\em purely by the stress tensor
itself} (unless body forces dominate). In this case, if one
director points along the major compression axis, then by symmetry any
others should lie at rightangles to it (Fig.~\ref{fig:jammed}1.(b)).
Applying
a similar argument to the intermediate axis leads to the ansatz that all
three directors lie along principal stress axes; this is perhaps the
simplest model in 3-d \cite{oslfoot}. 
(One version of this argument links force chains
with the fabric tensor \cite{kolymbas}, which is itself
typically coaxial with the stress \cite{radjai}.)  If so, Eq.(\ref{osl})
does not change form if an arbitrary isotropic pressure field $P$ is
added. With perpendicular directors as just described, Eq.(\ref{osl}) becomes
the ``fixed principle axes" ({\sc fpa}) model. We proposed this recently
to describe stress propagation in conical piles of
sand, constructed by pouring cohesionless grains from a point
source onto a rough rigid support. This model accounts
quite well for the forces measured experimentally beneath conical
sandpiles \cite{smidhuntley,nature,cargese}. More generally, Eq.(\ref{osl}) 
can be obtained by assuming a linear closure relation between the components
of the stress tensor \cite{nature}.

Although
the formation of dry granular aggregates under gravity is not normally
described in terms of jamming, it is closely related. Indeed, the
filling of silos and the motion of a piston in a cylinder filled with
grains exhibits jamming and stick-slip phenomena related directly to
formation of force chains in these geometries; see \cite{samchains}.
Hence fragile models of granular media must merit serious
consideration. They share some features with recent {\em hypoplastic}
models of such media
\cite{kolymbas}. Granular matter has, however, traditionally been
described by various forms of elastoplasticity, which we now compare
and contrast with the above idea of fragility.  

The characteristics evident in Eq.(\ref{osl})
are directly related to the {\em hyperbolic} nature of the
differential equations governing stress propagation in fragile
packings \cite{bcc,nature,cargese}.
With a zero-force boundary condition at the upper surface
of a pile \cite{cargese}, this gives a
well-posed problem: the forces acting at the base
follow uniquely from the body forces by integration: the Green function (in 2-d) comprises two rays connecting a
source to the base \cite{bcc,nature}. (Closely analogous
remarks apply in three dimensions.) The same does not hold \cite{roysoc}
for traditional elastoplastic
continuum models \cite{savagegoddard} whose equations
are (in simple cases) elliptic in elastic zones
and hyperbolic in plastic ones. In the sandpile, where an
elastic zone contacts the support,  the forces acting at the
base cannot then be found without specifying a
displacement field there. But such a displacement field has no
clear physical meaning for a sandpile created by pouring. To define
it, one requires a {\em reference state} in which the
stresses (gravity) are removed. Such a state is undefined, just as
it is undefined for a jammed colloid which, in the absence of
the applied shear stress, is a fluid.

This `elastic indeterminacy' of sandpiles has no
facile resolution \cite{roysoc}.  Suggestively,
the underlying problem (absence of a reference
state)  arises {\em precisely when}
elastoplasticity might give way to fragility: in
systems whose solidity arises because of the load
itself. Nonetheless, a crossover between fragile
and elastoplastic descriptions may exist, at
least in the context of very small incremental
loads (for which the reference state can be
defined in terms of a pre-existing,
gravitationally loaded pile). For example, one
{\em might} expect that, for poured sand, sound waves
of sufficiently small amplituded could propagate
normally (although in fact this is far from
obvious experimentally
\cite{nagelsound}). Likewise in our jammed colloid, {\em extremely
small} rotations of the principal axes might be accommodated by an
elastic, and not a plastic, mechanism.

We next show, for a specific example of a fragile
granular skeleton, that just such a crossover
can arise from a slight particle deformability.
We consider a highly idealized, 2-d rectangular
skeleton of rigid particles, Fig.\ref{fig:jammed}1.(b). In this material,
where the tangential compressive forces balance at each
node \cite{balledwards}, the shear stress must vanish across planes
parallel to $\bf n$ and
$\bf m$ (that is, $\sigma_{nm} = \sigma_{mn} = 0$).  For simplicity we
also assume that the ratio $\Lambda_1/\Lambda_2$ (and its inverse)
cannot exceed some constant $K$ (for example to avoid buckling of the
stress paths). This implies a Coulomb-like inequality, $|\sigma_{pq}|
\le \sigma_{qq}\,\tan\phi$, for all other orthogonal unit vector pairs
${\bf q},{\bf p}$; here $\tan\phi$ is a material constant such that $K =
(1-\sin\phi)/(1+\sin\phi)$.

Next a small degree of particle
deformability is introduced. This relaxes slightly the
colinearity requirement of forces along chains, because
the point contacts between particles are now flattened
(Fig.\ref{fig:paths}2.(b)). Clearly the ratio
$\epsilon$ of the maximum transverse load to the normal one will
vanish in some specified way (dependent on contact geometry)
with the mean particle deformation. The same ratio
$\epsilon$ defines, in effect, the maximum elastic angular
deviation of the force chains. The system can thus be described as
an (anisotropic) elastic body subject to a yield criterion of the
following form:
\be
|\sigma_{pq}| \le \sigma_{qq}\,\tan\Phi({\bf q \cdot n})
\label{coulomb}
\ee
where $\Phi(x)$ is a smooth function that is
small (of order $\epsilon$) in a narrow range (of order
$\epsilon$ wide) of orientations around $x=0$ (and $x=1$), but
close to $\phi$ outside this interval.

For finite $\epsilon$ this material will have mixed
elliptic/hyperbolic equations of the usual elastoplastic type.
But the resulting elastic and plastic zones must somehow arrange
themselves so as to obey the {\sc fpa} model to within terms that vanish
as $\epsilon \to 0$. If, in a sandpile,
$\epsilon$ is small but finite, then stresses will depend on the
detailed boundary conditions at the base of the pile, but only through
small corrections to the
leading ({\sc fpa}) result.
These deviations can accommodate an elastic
response to very small incremental loads (on a scale set by
$\epsilon$). But for the macroscopic stress pattern to differ
significantly from the hyperbolic prediction, one requires {\em
appreciable particle deformation}. When the mean stresses are
large enough to cause this ($\epsilon \simeq 1$), ``ordinary" elastic
or elastoplastic behavior will be recovered.  Conversely, the
fragile, hyperbolic limit emerges as {\em the limit of high
particle rigidity} for this simplified model skeleton.
Thus fragile models of granular or jammed matter, properly
interpreted, need not contradict (though equally they do not
require) an underlying elastoplastic description.

How valid are these ideas for granular media? The
existence of tenuous force-chain skeleton is hardly in doubt
\cite{dantu,coppersmith,thornton,kolymbas,radjai}.  Simulations of
frictional spheres show most of
the deviatoric stress to arise from force chains, while tangential
interparticle forces and
``spectator" contacts contribute mainly an isotropic pressure
\cite{thornton,radjai}. (Of course the specific geometry of
Fig.\ref{fig:jammed}1.(b) is grossly oversimplified: although the
force chains are anisotropic, they are not very straight and have
frequent collisions \cite{collisionfoot,thornton}.)
Are such skeletons actually fragile, as our models
suggest, or do they have an appreciable range of incremental elastic
response?  And, for sand under gravity, does a fragile or an
elastoplastic model better describe its
(nonperturbative) response to gravity itself?

We believe that granular media are often close to the fragile
limit. Firstly, we return to an earlier
argument: in systems whose solidity arises by a jamming process,
fragility may arise generically if the material arrests in states that
can only just support the applied load. This may be a reasonable
picture for sandpiles created by pouring. It could also apply to
unconsolidated dry grains in various other geometries.

Second, the probability distribution for
interparticle forces $p(f)$ does not vanish at zero force
\cite{coppersmith}. This is consistent with the idea that a slight,
incompatible change in load (relative scale $\delta/\bar f$ with $\bar
f$ the mean interparticle force) can induce a fraction $p(0)\delta$ of
contacts to switch from spectator type ($f\simeq 0$) to force-chain type
($f\simeq
\bar f$). The effect of such rearrangements would then be comparable
with the elastic response ($\bar f \to \bar f \pm
\delta$), and so formally destroy the elastic regime; we
expect this effect to be amplified by any long-range rearrangement
of force chains that a local contact change may induce \cite{staticavalanche}.
Third, simulations do indeed show
strong rearrangement of the skeleton under small
changes of compression axis; the skeleton is indeed
``self-organized" \cite{thornton,radjai}. There is also
evidence in some instances for internal cascades of
rearrangement \cite{staticavalanche,samchains}
in response to small disturbances. Although the latter is
strongly reminiscent of {\sc soc} \cite{soc}, our simplified models
show that fragility is a rather different concept which can arise, 
at least in principle, in regular geometries,
and without the self-similarity in the 
response to a small perturbation that characterizes {\sc soc}. In fact
{\sc soc}-like concepts underly recent discussions of dynamic
attractors in hypoplastic models \cite{kolymbas}, and
are not far removed from the (much older) critical state theories of
soil mechanics \cite{wood}.  The latter primarily address {\em
dilatancy}: the tendency of dense granular media to expand
upon shearing. Jamming can be viewed as the constant-volume
counterpart of this process.

We await further experimental guidance on the extent to which
granular materials are, in practice, fragile. Various experimental
tests of specific fragile models are suggested elsewhere
\cite{nature,cargese,staticavalanche}; these predict anomalies in
both correlation and response functions. More generally, the
negligibility of any incremental elastic range (as postulated in
fragile models for incompatible loads) can be probed by various
experiments including sound transmission. The latter do indeed show very
peculiar behaviour \cite{nagelsound}, possibly related to the fact that
the sound wave itself causes rearrangements. In addition, computer 
simulations should clarify the relationship between fragility and the extreme
nonlinearity which, for cohesionless dry sand, enters because tensile
contact forces are forbidden. Only when the probability density $p(0)$
of zero contact force becomes small can this be
safely ignored; and this might not arise before strong particle
deformation occurs ($\epsilon \simeq 1$).

More generally, other candidates for fragile matter
include jammed colloids, weak particulate gels, and flow-induced defect
textures in liquid crystals, all of which can self-organize so as to
support an applied stress.

We are grateful to P.~Evesque, P.G. de Gennes, G.~Gudehus, J.~Goddard,
J. Jenkins, D. Levine, S. Nagel, C.~Thornton, T.~A. Witten and 
especially S.~Edwards, for discussions.
This work was funded in part by EPSRC (UK) Grants GR/K56223 and GR/K76733.

Note added: we also refer the reader to C. Moukarzel, cond-mat/9803120, where
somewhat related ideas can be found.

\end{multicols}


\begin{thebibliography}{999}
\bibitem{oslfoot} A less symmetrical model allows all three
directors in Eq.(\ref{osl}) to be oblique. For example, in a
sandpile the construction history does not solely involve the stress
tensor, but also, e.g., the free surface orientation. See
\cite{nature}.
\bibitem{collisionfoot} Collisions between force chains
give the impression of a three-fold coordinated network in two
dimensions (see \cite{thornton}). Such a network is ``bulk
fragile" in contrast to Eq.\ref{osl}. However, individual collision
points might be micro-fragile, with changes in boundary load still
required to induce macroscopic rearrangement.
\end{thebibliography}
\end{document}